\documentclass[sigconf,nonacm]{acmart}
\pdfoutput=1
\usepackage{booktabs}
\usepackage{tabularx}
\usepackage{tikz}
\usetikzlibrary{arrows.meta,calc,fit,positioning,shapes.geometric}

\hypersetup{pdftitle={AI Infrastructure in Space: How Far Can We Go?}}
\begin{document}
\settopmatter{
  printacmref=false,
  printfolios=false
}

\pagestyle{plain}

\title{AI Infrastructure in Space: How Far Can We Go?}

\author{
Qing Li$^{*\dagger}$,
Qiyang Zhang$^{\ddagger}$,
Daliang Xu$^{*}$,
Tianze Huang$^{\ddagger}$,
Dingge Zhang$^{*}$,
Yihao Zhao$^{\ddagger}$,
Xiaolong Huang$^{\ddagger}$,
Jinfeng Wen$^{*}$,
Xiameng Hu$^{*}$,
Tao Qi$^{*}$,
Mengwei Xu$^{*}$,
Shangguang Wang$^{*}$,
Xuanzhe Liu$^{\ddagger}$
}

\affiliation{
$^{*}$Beijing University of Posts and Telecommunications\\
$^{\ddagger}$Peking University\\
$^{\dagger}$Corresponding author: Qing Li (qingli@bupt.edu.cn)\country{China}
}

\begin{abstract}
Satellites are becoming programmable computing platforms capable of running increasingly demanding AI workloads. This shift raises a systems problem: how can AI services remain deployable, manageable, and recoverable after launch when compute capacity, connectivity, energy, and thermal headroom vary over orbital time? This paper develops a systems vision for \emph{AI infrastructure in space}. We define it as the systems layer that manages AI capabilities across spacecraft, orbital networks, ground stations, and cloud backends, while treating orbital and physical state as part of the resource model. We synthesize relevant foundations from terrestrial AI infrastructure, satellite networking, and satellite edge computing, and examine the physical constraints that directly shape system design. We further ground this vision in three in-orbit case studies spanning the node, platform, and service levels. Telemetry from BUPT-1 satellite shows that usable compute capacity is bounded by thermal and energy envelopes. \textit{SateLight} on BUPT-2  satellite reduces application-update transmission latency by 56.54\% on average and up to 91.18\%, with 100\% update correctness. A stateful VLM serving case further shows that thermal interruptions make execution-state recovery a first-class systems problem. These observations motivate a research agenda for space-native resource management, lifecycle support, and sustained AI service across space and ground.
\end{abstract}

\maketitle
\section{Introduction}

AI progress increasingly depends not only on model architectures, but also on the infrastructure that makes models usable at scale. Modern AI systems require more than accelerators: they depend on memory and storage systems, high-speed interconnects, schedulers, deployment pipelines, and operational tools for serving, updating, and recovering models \cite{barroso2018datacenter,jouppi2017tpu,jouppi2023tpuv4,reddi2020mlperf,kwon2023vllm}. On Earth, these capabilities have been consolidated in datacenters, where AI infrastructure is built around grid power, active cooling, dense network fabrics, and human maintenance \cite{verma2015borg,hindman2011mesos,singh2015jupiter,poutievski2022jupiter}. As AI systems scale, however, this datacenter-centric model faces increasing pressure from power delivery, cooling capacity, land and water use, geographic latency, and the cost of moving data from its source to centralized compute.

Meanwhile, advances in launch systems, solar power, onboard accelerators, optical links, and programmable spacecraft are making increasingly capable computing systems possible in orbit. SpaceX has outlined large constellations of solar-powered orbital datacenters, Google's Project Suncatcher explores scalable machine-learning systems in space, and Starcloud is developing GPU-equipped orbital computing platforms. At the same time, AI has already been deployed onboard satellites for tasks such as cloud detection, telemetry analysis, link optimization, and autonomous planning \cite{giuffrida2022phisat,denby2020orbital,shen2024survey}. As these capabilities become more general and programmable, the systems problem extends beyond individual applications to the infrastructure that keeps them operational after launch. Models and their associated state must be deployed, updated, scheduled, monitored, and recovered across spacecraft, orbital links, ground stations, and cloud backends \cite{yu2022orca,kwon2023vllm,wen2025satelight}. We refer to this systems layer as \emph{AI infrastructure in space}.

Building this infrastructure requires a different resource model from terrestrial AI systems. Datacenter schedulers typically operate on a relatively stable pool of accelerators, supported by continuous power, active cooling, and high-speed networks \cite{verma2015borg,hindman2011mesos,jouppi2023tpuv4,gangidi2024rdma}. In orbit, usable compute capacity changes over time. Solar exposure and battery state determine the available energy budget, while limited heat rejection constrains how long intensive computation can be sustained. Connectivity follows orbital motion and contact opportunities rather than remaining continuously available. Satellites also differ in their energy and thermal states, network visibility, mission roles, and fault conditions, making them difficult to treat as interchangeable compute nodes. Because hardware cannot normally be repaired after launch, resource management must also account for reliability and recovery \cite{lee2013satellitepower,swanson2003nasa,corpino2015thermal,dodd2003basic,schwank2008radiation,fall2003dtn}. These properties make orbital compute a time-varying and non-fungible resource rather than a fixed pool of interchangeable accelerators.

A space-native AI infrastructure must turn this physical and orbital state into actionable information for system software. Runtimes should use platform telemetry and forecasts to guide workload placement, admission control, and service continuity \cite{denby2020orbital,denby2023kodan,lee2013satellitepower}. Orbital networks must support the movement of models, data, and execution state across satellites and ground systems, while the software stack must manage deployment, update, rollback, and recovery after launch \cite{yu2022orca,kwon2023vllm,wen2025satelight}. The objective is sustained and recoverable AI service over time rather than nominal compute capacity alone.

This paper presents a systems vision for AI infrastructure in space, grounded in both prior research and in-orbit experience. We first review the computing foundations most relevant to this problem, including terrestrial AI infrastructure, orbital networking, and satellite edge computing, and examine how their assumptions change in space. We then discuss the physical conditions that directly shape system design, with a focus on energy, thermal management, radiation, and spacecraft integration.
We further ground the discussion in three in-orbit case studies that span the node, platform, and service levels. Measurements from BUPT-1 show that usable COTS computing capacity is bounded by time-varying thermal and energy conditions. \textit{SateLight} on BUPT-2 demonstrates post-launch application update and rollback, reducing transmission latency by 56.54\% on average and up to 91.18\% while maintaining 100\% update correctness. Finally, a stateful VLM serving case shows how thermal interruptions turn long-running AI execution into a state-management and recovery problem. Based on these observations, we identify the key research challenges and outline a roadmap toward space-native AI infrastructure.

\section{What Is AI Infrastructure in Space?}

\subsection{Concept and Scope}

\subsubsection{Definition: From AI for Space to AI Infrastructure in Space}
We distinguish \emph{AI for space} from \emph{AI infrastructure in space}. AI for space uses models as mission tools, for example to detect clouds, flag telemetry anomalies, support autonomous planning or optimize links. AI infrastructure in space asks how such capabilities remain deployable, maintainable and reliable after launch. A spacecraft that runs one onboard model is AI-enabled. It becomes infrastructure only when that model can be deployed, updated, monitored, scheduled, recovered and coordinated with space and ground resources.

We define \emph{AI infrastructure in space} as a cross-layer system for deploying, orchestrating, operating, updating, evaluating and reliably managing AI capabilities across spaceborne compute, orbital networks and space-ground systems under physical and operational constraints. It comprises five coupled layers: a \emph{spaceborne compute layer} of processors, accelerators, memory and storage; an \emph{orbital network layer} of inter-satellite links, space-ground links, gateways and contact windows; a \emph{space-ground control plane} for deployment, monitoring, rollback, telemetry and scheduling; an \emph{AI lifecycle layer} for model versions, weights, adapters, checkpoints, KV caches, intermediate features and data products; and a \emph{physical-state interface} that exposes energy, thermal state, radiation risk, battery depth-of-discharge, attitude constraints, payload priority and mission safety margins.

The central question is whether AI can be operated as a managed service over time. Space AI infrastructure must therefore expose contact, power, heat, reliability and lifecycle state to the software stack, rather than burying them below it. 
Conceptually, the architecture forms a stack in which AI services sit above lifecycle and control-plane functions, with every layer coupled to communication constraints and the underlying physical state of the spacecraft.

\subsubsection{System Boundary}

The boundary of AI infrastructure in space is wider than the spacecraft. It includes onboard processors, accelerators, storage, runtimes and model-serving mechanisms, as well as the inter-satellite links, gateway access, ground-station schedules, model registries, update pipelines, mission-control software, cloud backends and digital twins that keep AI capabilities operational.

We draw this boundary around the lifecycle of an AI capability. Training or fine-tuning may occur on the ground; packaging and verification may occur in the cloud; upload is gated by contact windows; execution happens onboard; checkpoints and telemetry move across space and ground; and recovery may require rollback or fallback execution. Launch systems, propulsion, attitude control, orbital mechanics and payload hardware enter the model only when they affect AI operation. Orbit shapes contacts and energy cycles; attitude changes solar generation, antenna pointing and thermal exposure; payload priority constrains what AI may do. The transition is therefore from \emph{running AI in space} to \emph{operating AI as a managed space-ground service}.

\subsubsection{Key Difference from Terrestrial AI Infrastructure}
Terrestrial AI infrastructure is the closest design reference. Modern systems already rely on accelerators, high-speed datacenter fabrics, cluster managers, containerized deployment, model registries and serving runtimes \cite{barroso2018datacenter,verma2015borg,burns2016borg,moritz2018ray}. These mechanisms remain useful, but their assumptions change in orbit.

First, the compute node is not a stable server in a facility-managed pool. It is a spacecraft-bound resource whose usable capacity varies with solar exposure, eclipse, battery depth-of-discharge, thermal history, attitude, payload activity and mission rules. Onboard compute must therefore be modeled as physical-state-dependent capacity, not only as nominal TOPS or FLOPS.

Second, the network is not a persistent, low-latency datacenter fabric for collectives, storage access and rapid state movement \cite{jouppi2023tpuv4, gangidi2024rdma}. Satellites move; inter-satellite and space-ground links change; gateway access is limited by contact windows. The relevant resource is not only bandwidth, but when, for how long and with what reliability data can move. Delay-tolerant networking and contact-graph routing provide foundations \cite{fraire2021contact}, but AI infrastructure must also distinguish weights, checkpoints, KV caches, telemetry and sensor products.

Third, reliability cannot rely on physical maintenance. In orbit, radiation effects, software faults, link disruptions and thermal or power protection events require remote recovery, rollback, redundancy, checkpointing and degraded execution.

Finally, space nodes are not fungible. Satellites with identical processors may differ in orbit, contact opportunity, energy margin, thermal headroom, payload priority and fault risk. Model placement is therefore a joint compute, network, energy, thermal, reliability and mission-priority decision.

\subsection{Real-World Attempts Toward AI Infrastructure in Space}
Recent systems already point toward AI infrastructure in space, although none yet forms a complete stack. Together they demonstrate near-source inference, general-purpose spaceborne computing, programmable satellites, space-ground collaboration and early orbital data-centre visions. Their significance is the shift from fixed-function spacecraft toward programmable compute, network and service nodes.
\subsubsection{Onboard AI Applications}
Onboard AI is the most direct step toward near-source intelligence. Instead of downlinking all raw observations, a satellite can process data near the sensor and decide what to store, compress or transmit. ESA's $\Phi$sat-1, for example, used onboard AI to filter low-quality or cloud-covered Earth-observation images before downlink \cite{esa2020phisat,giuffrida2022phisat}. This reduced data volume and relieved scarce bandwidth.

Such systems remain closer to AI-enabled missions than to infrastructure: they typically bind one task, model and platform. The missing layer is lifetime update, monitoring, scheduling, recovery and coordination across space and ground.

\subsubsection{Spaceborne Computing Platforms}

Spaceborne computing platforms address the execution substrate. HPE's Spaceborne Computer and Spaceborne Computer-2, developed with NASA and deployed on the ISS, tested commercial high-performance computing in space and supported edge-computing experiments \cite{hpe2026spaceborne,nasa2025spaceborne}. They show that space systems can host general-purpose computing, not only fixed avionics or payload processors.

For AI infrastructure, this is necessary but insufficient. A platform becomes infrastructure only when its resources are managed as part of a service stack. Schedulers must see not only processor type and storage capacity, but also power, thermal headroom, contacts, reliability risk and mission priority.
\subsubsection{Cloud-Native and Programmable Satellites}
Programmable satellites shift the focus from hardware capability to operational flexibility. Containerized deployment, isolation, remote update, rollback, telemetry and ground-orbit DevOps make satellites behave less like fixed payloads and more like maintainable platforms. This matters for AI because models, adapters, preprocessing logic and serving runtimes evolve after launch.

Frameworks such as SateLight illustrate this direction by making application delivery, differential upload, onboard reconstruction and rollback first-class satellite functions \cite{wen2025satelight}. The implication is simple: evolving AI services require safe, observable software change.
\subsubsection{Space-Ground Collaborative AI Systems}
Near-term space AI infrastructure will be hybrid. Satellites are constrained by power, cooling, storage and maintainability; ground systems provide abundant compute and mature software ecosystems. Practical architectures therefore place latency-sensitive and bandwidth-saving tasks onboard, while training, validation, aggregation, long-term storage and fallback execution remain on the ground.

Cloud-integrated ground services already point in this direction. Microsoft Azure Orbital, for example, presents the ground segment as a cloud-connected platform for satellite communication, control, data processing and scalable operations \cite{microsoft2020azureorbital}. Such systems are not space AI infrastructure by themselves, but they provide the ground-side fabric for model deployment, telemetry, data products and recovery. Space-ground collaboration is therefore the default architecture, not an implementation detail.
\subsubsection{Orbital Data Centers and Space AI Service Visions}
A more ambitious direction is the orbital data centre. Axiom Space has described orbital data-centre nodes for space-based storage, processing and AI/ML services \cite{axiom2025odc}. Recent reports describe SpaceX plans for early orbital AI computing demonstrations before broader commercial deployment \cite{reuters2026spacexai}. These are visions or reported plans, not proof that hyperscale orbital AI datacentres are practical. Their importance is conceptual: they reframe space from a mission-specific environment into a possible substrate for selected compute, storage and AI services. Launch cost, power generation, thermal rejection, radiation tolerance, hardware replacement, ground egress, orbital safety and utilization remain open constraints. Even small hybrid systems make the infrastructure question unavoidable.
\subsection{From Isolated Attempts to Infrastructure}
The attempts reviewed above show a clear trend: satellites are moving from fixed-function mission assets toward programmable compute, network and service nodes. Onboard AI demonstrates near-source inference; spaceborne computing platforms validate general-purpose processing; programmable satellites enable post-launch software evolution; space-ground services connect orbital assets with cloud systems; and orbital data-centre visions point to longer-term infrastructure \cite{esa2020phisat,giuffrida2022phisat,hpe2026spaceborne,nasa2025spaceborne,wen2025satelight,microsoft2020azureorbital,axiom2025odc,reuters2026spacexai}.
\subsubsection{What Has Been Demonstrated}
Existing systems have demonstrated key pieces of the stack: onboard inference can reduce downlink demand; commercial compute can operate in space; software can be remotely deployed, updated, monitored and rolled back; and ground-cloud platforms can support communication, data processing and operations. AI capabilities can therefore exist in orbit and connect to ground workflows.
\subsubsection{What Is Still Missing}
What remains missing is the infrastructure layer that makes these capabilities reusable, schedulable and reliable across missions and platforms. Current systems are still tied to specific tasks, satellites, models or demonstrations. They lack common resource abstractions, model and state lifecycle management, and cross-layer orchestration across compute, contact windows, energy, thermal state, reliability and mission priority.
\subsubsection{Key Takeaway}
Existing systems demonstrate pieces of the stack, but not yet a space-native infrastructure layer. The next step is to move from \emph{running AI in space} to \emph{operating AI over space-ground systems}: satellites, orbital links, ground stations, model artifacts, telemetry and physical constraints must be managed together.


\section{Computing Foundations and Related Work}

\subsection{Terrestrial AI Infrastructure}

We use terrestrial AI infrastructure as the baseline whose mechanisms space systems may inherit, and whose assumptions they must revise. We include work that defines datacenter-scale AI compute, governs shared heterogeneous workloads or distributes training and inference across accelerators. The critical assumptions are abundant power, active cooling, fungible nodes and an always-on lossless fabric. None survives unchanged in orbit.

\subsubsection{AI Datacenter Architecture}
Modern AI datacenters centre on GPUs and TPUs as the unit of compute: their datapaths, high-bandwidth memory and interconnect define per-node capability~\cite{jouppi2017tpu,jouppi2023tpuv4,choquette2021a100}, which is procured and reported through benchmarks~\cite{reddi2020mlperf,mattson2020mlperf}. Because training throughput is bounded by collective communication, the network becomes the computer; this has driven non-blocking fabrics, optically reconfigurable topologies~\cite{singh2015jupiter,poutievski2022jupiter,jouppi2023tpuv4}, and GPU fabrics tuned for LLM traffic and fault tolerance~\cite{qian2024hpn,gangidi2024rdma}. The cluster is the management unit, since warehouse-scale operation relies on cluster managers that pool and allocate machines as one computer~\cite{verma2015borg,tirmazi2020borg,hindman2011mesos,burns2016borg,barroso2018datacenter}. Power and cooling are treated as facility services: terrestrial designs assume grid power, water or air cooling and microsecond-scale lossless interconnect, and power is managed through over-subscription rather than scarcity~\cite{barroso2017killer,patel2024power,patel2023polca}. Finally, at frontier-scale operation, public reports of ten-thousand-accelerator clusters show that stragglers, failures and imbalance, not peak FLOPS alone, limit operation~\cite{jiang2024megascale,hu2024characterization,dubey2024llama3,deepseekai2024v3}.

\textit{Implication:} the terrestrial stack treats compute, network, power and cooling as abundant, homogeneous and continuously available. Section~4 revisits these assumptions under orbital physics.

\subsubsection{Resource Management and Scheduling}
GPU schedulers exploit the iterative structure of training to reduce completion time and raise utilization through time-slicing, migration and information-agnostic policies~\cite{xiao2018gandiva,gu2019tiresias,mahajan2020themis,peng2018optimus}. Later schedulers address heterogeneity and elasticity, co-adapting allocation with batch size or goodput and reasoning about heterogeneous accelerators and elastic jobs~\cite{qiao2021pollux,narayanan2020gavel,subramanya2023sia,zheng2023shockwave,gu2023elasticflow,hwang2021afs,li2023lyra}. Fine-grained sharing lets multi-tenant clusters provide guarantees and GPU sharing through cell-based allocation, co-execution, fast context switching and memory sharing~\cite{zhao2020hived,xiao2020antman,bai2020pipeswitch,lim2021zico,mohan2022synergy}. Scheduler design is also trace-grounded, calibrated against production GPU traces~\cite{jeon2019analysis,weng2022mlaas,hu2024characterization}, while general-purpose frameworks support elastic execution~\cite{moritz2018ray}.

\textit{Implication:} these policies assume interchangeable, always-available accelerators that can be packed, preempted, and migrated at will. LEO nodes violate this assumption because energy and thermal state gate availability.

\subsubsection{Distributed Training and Inference}
Training parallelism scales beyond one device by composing data, tensor, pipeline and expert parallelism, either automatically or by hand tuning~\cite{shoeybi2019megatron,narayanan2021megatron,huang2019gpipe,narayanan2019pipedream,rajbhandari2020zero,zheng2022alpa,jia2019flexflow,xu2021gspmd,rasley2020deepspeed,zhao2023fsdp,barham2022pathways}. Sparsity and mixtures of experts use conditional computation to expand parameter count at near-constant FLOPs, but increase all-to-all communication and routing pressure~\cite{lepikhin2021gshard,fedus2022switch,rajbhandari2022deepspeedmoe,deepseekai2024v3}. Because long jobs fail often, fault tolerance makes checkpoint/restart, in-memory recovery, redundant recovery and preemption-aware reconfiguration central concerns~\cite{mohan2021checkfreq,eisenman2022checknrun,wang2023gemini,thorpe2023bamboo,jang2023oobleck,athlur2022varuna}. For inference and serving, generative serving introduced iteration-level batching, paged KV-cache memory, prefill/decode disaggregation, statistical multiplexing, and KV-centric or migration-based scheduling~\cite{yu2022orca,kwon2023vllm,zhong2024distserve,agrawal2024sarathi,patel2024splitwise,qin2025mooncake,li2023alpaserve,sun2024llumnix,fu2024serverlessllm,dao2022flashattention,dao2023flashattention2}, building on model-serving systems that set latency-SLO baselines~\cite{crankshaw2017clipper,gujarati2020clockwork,romero2021infaas}. To sustain efficiency under constraint, offloading, speculative decoding and optimized engines stretch fixed hardware~\cite{sheng2023flexgen,leviathan2023speculative,aminabadi2022deepspeedinference}, while model reports document full training and serving stacks~\cite{dubey2024llama3,deepseekai2024v3,bai2023qwen,openai2023gpt4}.

\textit{Implication:} both training and serving assume synchronized collectives over high-bandwidth, lossless and always-on interconnects among co-located accelerators. A contact-window-mediated orbital fabric cannot provide this property.

\subsubsection{Assumptions That Break in Space}
The terrestrial stack is effective, but four assumptions do not survive the move to orbit. First, abundant, steady power gives way to harvested, cyclic energy: datacenter scheduling treats power as oversubscribable~\cite{patel2024power,patel2023polca}, whereas in orbit usable compute is gated by solar exposure, eclipse and battery depth-of-discharge (Sections~4.1 and~5.2). Second, active cooling gives way to radiative-only heat rejection, as air and water cooling~\cite{barroso2017killer} yield to conduction and radiation, so that thermal headroom, not raw FLOPS, limits sustained serving (Sections~4.2 and~5.4). Third, fungible nodes give way to non-fungible nodes: schedulers pack, preempt and migrate across interchangeable accelerators~\cite{xiao2018gandiva,qiao2021pollux,sun2024llumnix}, but satellites differ in orbit, energy state and contact opportunity (Sections~4.3 and~4.4). Fourth, the lossless fabric gives way to scheduled links, since distributed training and serving assume high-bandwidth lossless collectives~\cite{narayanan2021megatron,rajbhandari2020zero,kwon2023vllm} while orbital interconnect is contact-window-mediated and energy-coupled (Section~3.2).

\noindent\textbf{Inherited:} accelerator abstractions, parallelism strategies, cluster schedulers and lifecycle/serving mechanisms. \textbf{Broken:} the resource model, the failure model and the connectivity model.
\textit{Takeaway:} space AI infrastructure should reuse terrestrial mechanisms while replacing their assumptions with physically grounded, predictable-but-intermittent models (Sections~4--6). Table~\ref{tab:terrestrial-vs-space} summarizes the replacement.

\begin{table*}[t]
\caption{Terrestrial AI infrastructure assumptions and space-native replacements.}
\label{tab:terrestrial-vs-space}
\centering
\small
\begin{tabularx}{\textwidth}{p{0.18\textwidth} X X}
\toprule
\textbf{Dimension} & \textbf{Terrestrial AI infrastructure} & \textbf{Space-native AI infrastructure} \\
\midrule
Compute node & Server or accelerator tray inside a repairable, facility-managed pool. & Spacecraft-bound compute resource whose usable capacity varies with orbit, battery, thermal state, payload priority, and fault risk. \\
Power and cooling & Grid-backed power and active air or liquid cooling treated mainly as facility constraints. & Harvested solar energy, eclipse, depth-of-discharge, conduction, and radiative heat rejection directly gate service time. \\
Network fabric & Always-on, low-latency, high-bandwidth fabric for collectives, storage, and migration. & Scheduled and intermittent contact graph with link opportunities that depend on orbit, pointing, energy, and gateway visibility. \\
Scheduler model & Pools of mostly fungible devices optimized for utilization, fairness, and SLOs. & Non-fungible orbital nodes scheduled over time-varying contact, energy, thermal, reliability, and mission-priority envelopes. \\
Lifecycle state & Models, containers, checkpoints, and serving state move through stable datacenter networks. & Weights, deltas, adapters, checkpoints, KV caches, and telemetry must be staged across constrained uplinks, onboard storage, and rollback paths. \\
Failure and recovery & Failed hardware can often be repaired, replaced, rebooted, or masked by nearby capacity. & Radiation, thermal protection, power events, and link loss require remote recovery, degraded modes, redundancy, and auditable rollback. \\
Evaluation & Benchmarks emphasize throughput, latency, utilization, and cost under stable infrastructure. & Benchmarks must include per-orbit useful work, energy and thermal margins, contact usage, recovery behavior, and mission safety. \\
\bottomrule
\end{tabularx}
\end{table*}

\subsection{Satellite and Orbital Networking}
 
We treat orbital networking as the interconnect and I/O fabric of space AI infrastructure, not as a stand-alone communication service. The relevant work constrains where AI workloads can be placed, when nodes are reachable, how fast models and state can move, and whether that movement can be planned.
 
\subsubsection{The Orbital Network Substrate}
The substrate is layered by orbit regime. GEO, MEO and LEO trade altitude against latency, coverage and per-node resources. LEO mega-constellations, including Starlink, OneWeb, Kuiper, Guowang and Qianfan, are plausible near-term homes for spaceborne compute~\cite{delportillo2019technical}. Within a constellation, satellites connect through radio-frequency or optical inter-satellite links, often in a \textit{+Grid} motif of two intra-plane and two inter-plane neighbours. Optical links already reach the ${\sim}100$~Gbps class~\cite{chaudhry2021laser}. Their system-visible properties are link establishment delay, outage probability and topology reconfiguration time.

The space--ground segment adds user links and feeder or gateway links. These are increasingly optical, weather-coupled, offered as ground-station-as-a-service and studied as shared distributed resources~\cite{vasisht2020distributed}. The resulting fabric is asymmetric. In-space mesh capacity scales with constellation size, whereas ground egress is capped by gateways, spectrum and contact time. Data, and increasingly computation, therefore accumulate in orbit~\cite{bhattacherjee2020inorbit}.

\textit{Implication:} the near-term fabric is globally distributed and mesh-rich, but egress-constrained, mechanically steered, energy-coupled and never fully active at once.

\subsubsection{Predictable Topology Dynamics}
The distinctive property of orbital networking is predictable motion. Connectivity varies continuously, but ephemerides make topology computable hours to days ahead, unlike static datacenter fabrics or stochastic terrestrial MANETs. Snapshot and virtual-topology models, virtual nodes~\cite{werner1997dynamic,
ekici2001distributed}, contact graphs, contact-plan design and time-expanded graph optimization~\cite{fraire2015design, ron2025timedependent} turn a moving constellation into analyzable graph sequences.

Topology is also a design variable. ISL motifs set latency and throughput envelopes~\cite{bhattacherjee2018gearing, bhattacherjee2019network}, and whether routing through space beats ground relay remains contested~\cite{handley2018delay, handley2019using}. Predictability is imperfect: operational constellations are actively managed and reconfigured~\cite{li2023networking}, while user and gateway handovers create minute-scale path churn.

\textit{Implication:} predictability is the main asset orbital networks have that terrestrial fabrics lack. Current AI runtimes, however, expose no interface to consume it, a gap we revisit in the runtime challenges.

\subsubsection{Moving Bits Under Intermittency}
Given a known but time-varying topology, routing must trade stability against optimality as links appear and vanish. Snapshot shortest-path, geographic and source-routing schemes address changing graphs. Under intermittency, however, movement must be planned. Delay/disruption-tolerant networking, store-carry-forward, the Bundle Protocol and Contact Graph Routing~\cite{fall2003dtn, rfc9171, araniti2015contact} therefore bear directly on transferring weights and checkpoints. Egress depends on contact-window and ground-station scheduling, which is now optimized for latency and throughput across distributed ground networks~\cite{vasisht2021l2d2, tao2023transmitting}, and is coupled to onboard data reduction.

At the transport layer, TCP and QUIC degrade over LEO paths because handovers inject loss and jitter. Link-layer-informed adaptation~\cite{cao2023satcp} and measurements of operational constellations~\cite{michel2022first,
mohan2024multifaceted} matter because they bound how fast weights, checkpoints and state move within contact windows. This understanding depends on emulators and testbeds such as Hypatia, StarPerf, StarryNet, Celestial, OpenSN and LEOCraft~\cite{kassing2020hypatia, lai2020starperf,
lai2023starrynet, pfandzelter2022celestial, lu2025opensn, basak2025leocraft}; RHONE~\cite{wang2025rhone} extends this methodology toward space computing networks.

\textit{Implication:} this work explains where and when bits can flow, but treats traffic as opaque bytes. It does not distinguish weights, KV caches, gradients and telemetry, nor co-schedule them with compute, energy or thermal state.

\subsubsection{From Communication to Computing Interconnect}
Taken together, four assumptions of terrestrial AI fabrics break in orbit. Static topology becomes predictable dynamics, so contact windows should be first-class scheduling resources. Microsecond, lossless RDMA becomes millisecond-to-second intermittent paths, so collectives require redesign rather than porting. Commodity bandwidth becomes a scheduled, energy-coupled resource because link activation competes with payload and thermal budgets. Any-to-any reachability becomes contact-window-mediated reachability. Existing work supplies the lower layer: topology and contact formalisms, dynamic-graph routing, DTN scheduling, downlink scheduling and emulation tools. The missing layer is AI-specific: traffic classes for model dissemination, state synchronization, checkpointing and telemetry; deadline- and contact-aware bulk transfer; predicted network state exposed to AI runtimes; and joint compute--network scheduling.

\textit{Takeaway:} the orbital network should be recast as the interconnect and I/O layer of a space-native computing system. The contact window, not the link, is the basic allocation unit.

\subsection{Satellite Edge Computing and Onboard Intelligence}

Satellite edge computing moves computation toward the data rather than the data toward computation. Instead of downlinking raw observations for processing on the ground, satellites can run inference in orbit and return only compact, higher-value products. This shift raises a set of coupled questions---what to compute onboard, how to divide workloads across onboard processors, ground-station edge nodes and terrestrial clouds, how satellites should cooperate, and what runtime and evaluation support such systems demand---that we examine in turn below.

\subsubsection{Data Filtering, Compression, and Selective Downlink}

Modern Earth observation missions face a data deluge: sensing capability grows faster than intermittent space--ground links. This is acute for high-resolution optical, hyperspectral, SAR and video missions, where raw observations can exceed contact-window capacity. Onboard processing moves the first stage of selection into orbit, allowing satellites to discard low-value scenes, prioritize urgent events or produce compact semantic products before downlink. $\Phi$-Sat-1 and CloudScout demonstrate neural-network cloud screening for hyperspectral imagery~\cite{giuffrida2020cloudscout,giuffrida2022phisat}. OPS-SAT shows reconfigurable in-orbit image classification, clustering, cloud detection and telemetry anomaly detection~\cite{labreche2022smartcam,kacker2022opssat,meoni2024opssat,ruszczak2025opssatad}. HYPSO-1 extends this direction with in-orbit sea--land--cloud segmentation on real hyperspectral data~\cite{langer2023hypso,justo2023hypso,justo2025hypso}. Together, these systems turn satellites from passive data sources into active information filters.

This shift changes how data reduction should be evaluated. Compression is usually judged by reconstruction fidelity under a bit budget. Onboard intelligence should instead be judged by mission utility under communication, compute, storage and energy constraints. Outputs may be cloud masks, object chips, segmentation maps, embeddings, confidence scores or event alerts. They reduce downlink volume, but introduce model-dependent decision risk. Selective downlink should therefore be evaluated by bandwidth savings, downstream accuracy, timeliness, uncertainty calibration, auditability and the cost of discarding observations that later prove important~\cite{denby2023kodan,soret2024semantic}.

\subsubsection{Compute-Network Joint Scheduling and Resource Allocation}
Satellite edge computing operates under strict limits on processing, memory, storage, energy, thermal dissipation and contact opportunities. Local inference can reduce downlink demand and latency, but it competes with sensing, storage and communication for spacecraft resources. Scheduling must therefore decide where a task runs, when its data move and whether intermediate results are stored, forwarded or discarded. This has motivated studies of computation offloading, resource allocation, service placement and computing-aware routing across satellite--terrestrial networks~\cite{zhang2019smec,pfandzelter2022qos,rossi2025resource,cao2023computing}.

Existing formulations optimize task offloading, bandwidth, power, compute, routing or service caching. Representative work studies cooperative offloading, energy-aware placement, QoS-aware resource placement and end-to-end offloading with multidimensional allocation~\cite{wang2020joint,song2021energy,tang2021computation,qu2024computation}. For AI-driven Earth observation, delay or energy alone is insufficient. The scheduler must distinguish raw observations, compressed products, semantic masks, embeddings, alerts and model updates because these objects differ in urgency, size, uncertainty and downstream value. Kodan and Serval highlight this semantic dimension: computation should maximize delivered information value under latency and resource constraints~\cite{denby2023kodan,tao2024serval}.

\subsubsection{Multi-Satellite Collaboration and Distributed Intelligence}
Multi-satellite collaboration follows from the distributed nature of constellation sensing. One satellite has limited revisit frequency, partial coverage and constrained onboard resources. Collaboration can exchange compact features or alerts, partition inputs, split DNNs across satellites or space--ground nodes, or update models through federated learning without centralizing raw observations~\cite{xu2023coinleo,guan2024collaborative,zhang2024edijp,wu2023fedgsm,shi2024satellitefeel}. Unlike terrestrial edge systems, these mechanisms must handle intermittent links, non-IID observations, heterogeneous resources and stale model or feature updates.

The need grows as satellite AI moves toward larger models. Foundation models can improve retrieval, segmentation, change detection and multimodal reasoning, but full onboard deployment is limited by memory, energy and thermal constraints. A practical direction is distributed execution across constellations and the ground segment~\cite{shi2025large}.

\subsubsection{Runtime, Reliability, and Evaluation Infrastructure}

Operational satellite edge AI needs runtime support beyond mission scripts. A reusable runtime should expose workload structure, data-product type, deadline, priority, resource demand, confidence and fallback behavior. Schedulers can then choose onboard execution, ground-ingress execution or cloud execution. Reliability is part of the same interface because onboard AI may discard data or trigger actions before ground verification. Operational systems therefore need versioning, rollback, health monitoring, watchdogs, confidence-aware transmission and safe degradation under missed contacts, throttling, low energy or thermal constraints.

Evaluation infrastructure remains immature. In-orbit experiments are expensive, safety-constrained and hard to reproduce. Conventional simulators often miss spacecraft resource dynamics and AI workload behaviour. StarryNet emulates integrated space--terrestrial networks, and RHONE adds satellite-level power, thermal, orbit, network and computation dynamics~\cite{lai2023starrynet,wang2025rhone}. Satellite edge AI benchmarks should also report model latency, memory footprint, confidence distributions, intermediate data sizes, update cost, energy use and mission utility. HYPSO-1 sea--land--cloud imagery and OPSSAT-AD telemetry provide starting points~\cite{justo2023hypso,ruszczak2025opssatad}, but integrated benchmarks spanning sensor data, workload DAGs, orbital traces, contacts, hardware profiles and mission-risk metrics are still missing.

\subsection{Synthesis: Building Blocks, but Not Yet Infrastructure}

\subsubsection{What Existing CS Work Provides}
The computing and networking literature already provides reusable mechanisms. Terrestrial AI infrastructure contributes accelerator-centric abstractions, cluster scheduling, distributed training, model serving, checkpointing and deployment. Orbital networking contributes contact prediction, delay-tolerant transfer, routing over time-varying graphs, ground-station scheduling and emulation tools. Satellite edge computing contributes near-source filtering, selective downlink, collaborative inference and mission-aware data products. The pieces exist: compute can be shared, models and state can move, workloads can be scheduled and network dynamics can be predicted.

\subsubsection{What Remains Fragmented}
The problem is composition. AI systems assume stable datacenter power, cooling and connectivity. Orbital networking treats traffic mostly as bytes, not as weights, checkpoints, KV caches, features or telemetry. Satellite edge systems often optimize one mission pipeline rather than a reusable service layer. Existing work therefore lacks a resource model that jointly exposes contact windows, compute capacity, storage pressure, energy margin, thermal headroom, reliability risk and mission priority. Lifecycle mechanisms are also fragmented: update, rollback, checkpointing, evaluation and recovery are separate functions rather than one operating substrate.

\subsubsection{Toward a Space-Native AI Infrastructure Stack}
The next step is not to port a datacenter stack into orbit. It is to rebuild the infrastructure interface around predictable physical and network state. Spaceborne compute should be a time-varying service envelope, not a fixed accelerator pool. Orbital links should be the interconnect and I/O layer for model artifacts, state, telemetry and data products. The control plane should coordinate deployment, scheduling, update, monitoring, rollback and evaluation across space and ground. Energy, thermal state, radiation risk, contact forecasts and mission safety must be visible to the AI runtime. The next section explains why these constraints are the resource model, not implementation details.

\section{Physical Foundations of Space AI Infrastructure}
\label{sec:physical-foundations}

\subsection{Energy as a First-Class Resource}
\label{subsec:energy-first-class}

\noindent\textbf{Energy determines service continuity.}
Energy is not only a cost in orbit; it is a condition for service continuity. An AI node must power payloads, avionics, attitude control, storage, radios, heaters, processors, and accelerators while preserving battery and bus margins across sunlight and eclipse. The provisioning question is therefore not peak TOPS under laboratory power, but whether inference, planning, compression, or learning can be sustained over an orbital horizon without violating platform reserves. Orbital edge-computing work treats satellites as distributed computers whose usable compute depends on onboard resource envelopes, contacts, and mission timing \cite{denby2020orbital,denby2023kodan}. COTS satellite measurements and space--air--ground surveys likewise show that deployable AI is bounded by coupled compute, energy, communication, and environmental constraints \cite{xing2024cots,shen2024survey}. Capacity should therefore be reported in energy-feasible units, such as useful inferences per orbit, observations filtered per watt-hour, or latency under specified power reserves.

\noindent\textbf{Orbit and attitude shape solar supply.}
Solar supply is predictable only after orbital geometry, eclipse timing, array orientation, temperature, degradation, conversion efficiency, and attitude plans are considered. Orbital period and beta angle set sunlight and eclipse duration, while imaging, antenna pointing, formation control, and momentum management can reduce incident power through cosine loss or self-shadowing. Power-system and panel-orientation studies show how generation, storage, and load profiles vary with these choices \cite{lee2013satellitepower,anigstein1998solar}. Energy-harvesting systems provide the relevant abstraction: the runtime should exploit a forecastable energy-arrival process instead of assuming a stationary budget \cite{kansal2007power,hester2017timely}. Ephemerides, attitude timelines, payload schedules, and contacts should therefore become a time-varying power envelope that the AI scheduler can query.

\noindent\textbf{Battery state bounds usable compute.}
Battery capacity is also not directly convertible into AI work. Admissible load depends on state of charge, depth of discharge, discharge rate, cell temperature, aging, imbalance, conversion loss, and mission safety margins. Battery-aware scheduling shows that workloads with the same energy demand can have different lifetimes when their discharge profiles differ \cite{luo2001batteryaware,rakhmatov2003energy}. Cyber-physical power management and intermittent-system designs further show that admission control must preserve protected functions and recovery reserves under supply uncertainty \cite{kim2016offline,colin2018capybara}. A satellite runtime should enforce state-of-charge limits, temperature derating, C-rate limits, aging uncertainty, and safe-mode reserves, scaling or rejecting AI work before battery health or bus voltage becomes critical.

\noindent\textbf{AI accelerators create power peaks.}
Specialized accelerators improve energy efficiency through reduced precision, data reuse, sparse execution, and domain-specific dataflows \cite{jouppi2017tpu,han2016eie,chen2017eyeriss}. Yet low energy per inference does not guarantee benign instantaneous power. Dense tensor phases, memory bursts, wake-up, regulator transients, and overlap with radio transmission can create peaks far above the average. Computational sprinting shows the value of briefly exceeding sustainable power or thermal levels, but the same tactic is risky when the accelerator shares a regulated spacecraft bus with payloads, actuators, heaters, storage, and radios \cite{raghavan2012sprinting}. Qualification should therefore report peak power, ramp rate, rail demand, memory and I/O power, and behavior under caps, and system software should coordinate accelerator phases with DVFS, radio duty cycles, and payload operations.

\noindent\textbf{The full pipeline shares one energy budget.}
An AI output is produced by a pipeline, not a processor alone. The spacecraft senses, moves and stores data, preprocesses, infers, compresses or prioritizes outputs, schedules downlink, and may change thermal-control behavior. Near-real-time Earth-observation systems show that onboard and ground processing should be chosen jointly, because in-orbit filtering reduces latency and downlink demand but consumes onboard energy \cite{tao2024serval}. Contact-aware transmission work shows that communication energy and latency depend on temporal and spatial opportunities \cite{tao2023transmitting}. Satellite-computing studies that include real-time, energy, and temperature constraints show that execution, communication, and thermal behavior cannot be optimized separately \cite{li2024realtime,tang2024joint}. The scheduler should maximize mission value per constrained joule across the full sensing-to-delivery path.

\noindent\textbf{Runtime must be energy-aware.}
The systems requirement is an AI runtime that exposes energy state to admission, scheduling, and degradation. It should combine ephemeris-derived harvesting forecasts, attitude and payload schedules, battery health, rail-level telemetry, and model profiles for latency, accuracy, memory, average power, and peak power. Adaptive inference and constellation emulation show how accuracy, latency, and energy can be evaluated against orbital dynamics before launch \cite{liu2025spaceexit,wang2025rhone}. Intermittent inference and time-sensitive intermittent learning show how execution can be split into restartable units when harvested energy is unstable \cite{gobieski2019sonic,islam2020zygarde}. The runtime should reserve non-preemptible energy for critical subsystems, choose model variants over one or more orbits, and degrade explicitly when margins shrink.

\subsection{Thermal Management as the Hidden Bottleneck}
\label{subsec:thermal-hidden-bottleneck}

\noindent\textbf{Vacuum makes cooling radiative.}
Thermal management is hidden but often binding because near-vacuum removes the convective heat sink used by terrestrial servers. Heat from an AI die must conduct through the package, board, structural interfaces, and heat-transport hardware before being rejected by radiation. Space thermal control therefore relies on conduction paths, radiators, coatings, heat pipes, capillary devices, phase-change materials, and view factors rather than fans and ambient airflow \cite{swanson2003nasa,figus2003capillary,hoa2003heatpipes}. Small-satellite design further shows that the body, payload placement, external surfaces, and orbital heat inputs form one coupled network \cite{corpino2015thermal}. Compute planning must therefore reason about heat generation and heat rejection, not only electrical power.

\noindent\textbf{AI chips create local hotspots.}
High-performance AI processors can create local hotspots even when the average spacecraft temperature is safe. Tensor units, memory controllers, regulators, stacked memories, and I/O PHYs concentrate heat, affecting timing, leakage, aging, package stress, and nearby payloads. Dynamic thermal-management work established that microarchitectural activity must respond to temperature, not average power alone \cite{brooks2001dtm}. HotSpot and multicore thermal-management studies link architectural power maps to compact thermal models and show that placement and migration can reduce hotspots \cite{skadron2003temp,huang2006hotspot,donald2006multicore}. AI accelerators should therefore be characterized by spatial and temporal power maps, and schedulers should know which kernels heat which physical regions.

\noindent\textbf{Radiators and layout bound heat rejection.}
Heat rejection is constrained by radiator area and orientation, surface optical properties, interfaces, structural layout, and blockage by arrays, antennas, baffles, or payload apertures. Deployable radiator and panel studies show that increasing rejection capacity is a mechanical and geometric problem \cite{bulut2015panel}. CubeSat and pico-satellite analyses show that limited surface area, orientation, material choice, and packaging dominate small-spacecraft temperatures \cite{kovacs2018smog,bonnici2019pico}. Radiator work further emphasizes integration with heat transport and structure \cite{torres2014satellite}. Sustained AI compute therefore depends on processor placement, heat-path quality, radiator view, and conflicts with power generation or communication.

\noindent\textbf{Orbits impose thermal cycles.}
The orbital thermal environment is periodic and history-dependent. Solar flux, Earth albedo, planetary infrared radiation, internal dissipation, and radiative cooling change over each orbit, with sharp transitions at eclipse entry and exit. Studies of small satellites and passive control show that these transitions create repeating temperature swings shaped by thermal capacitance, coatings, geometry, and internal power schedules \cite{anh2016radiation,escobar2016evolutionary,elhefnawy2022passive,corpino2015thermal}. AI execution interacts with these cycles: an inference burst before eclipse may leave stored heat with little rejection opportunity, while the same burst during a favorable radiative period may be safe. Scheduling should therefore use both current temperature and forecasted heat-rejection opportunity.

\noindent\textbf{Sustained compute is thermally bounded.}
Peak TOPS or FLOPS does not define deployable AI capacity when heat cannot be continuously removed. Dark-silicon results show that power and thermal limits can leave much silicon unavailable for simultaneous activation \cite{esmaeilzadeh2011dark}. Work on Heat-and-Run, thermal herding, and multiobjective thermal control similarly shows that throughput depends on duty cycle, thermal capacitance, spatial heat distribution, and cooling constraints \cite{gomaa2004heatrun,puttaswamy2007thermal,sabry2011multiobjective}. In orbit, the heat sink is finite, radiative, orientation-dependent, and shared. Useful metrics should therefore report sustained mission throughput under thermal margins, transient warm-up, steady-state temperature, cooldown time, and allowed duty cycle.

\noindent\textbf{Runtime must enforce thermal safety.}
A thermal-aware runtime should act before safety limits are reached. Proactive temperature balancing and hierarchical thermal management show that future thermal states can be shaped by present scheduling and by coordination between local controls and global placement \cite{coskun2008proactive,zanini2011hierarchical}. Temperature-aware and real-time satellite-computing work connects this principle to orbital workloads by considering deadlines, energy, and temperature together \cite{wang2026temperature,li2024realtime}. A space AI runtime should maintain a reduced-order thermal model calibrated by sensors, map kernels to heat profiles, forecast sunlight and eclipse transitions, and choose among placement, DVFS, batching, model simplification, deferral, and safe degradation.

\subsection{Radiation, Reliability, and Non-Maintainability}
\label{subsec:radiation-reliability-nonmaintainability}

\noindent\textbf{Radiation causes faults and degradation.}
Space radiation affects AI infrastructure through transient faults and cumulative degradation. Single-event upsets, transients, and latchup can corrupt state or create destructive current paths, while total ionizing dose and displacement damage degrade devices over mission lifetime. Foundational radiation-effects work shows that susceptibility depends on particle type, energy, device geometry, process technology, shielding, and operating voltage \cite{dodd2003basic,schwank2008radiation}. Soft-error research shows that hardware faults become system failures through architectural and software masking chains, so vulnerability is a stack property rather than a device property alone \cite{mukherjee2005soft}. SRAM-FPGA studies further show that configuration state can dominate reconfigurable-platform vulnerability \cite{ferlet2013set}. Space AI must therefore bound the probability that corrupted weights, activations, metadata, storage, buffers, or control decisions reach mission outputs.

\noindent\textbf{COTS reliability is uncertain.}
COTS processors and accelerators are attractive because they provide modern process nodes, high arithmetic density, mature software, optimized compilers, and fast deployment. Their weakness is radiation behavior that is often uncertain, workload-dependent, and insufficiently characterized for mission assurance. Neural-network radiation studies show that learning workloads may tolerate some bit errors yet fail sharply when faults strike critical weights, activations, or control structures \cite{rech2024ann,libano2020quantization}. Proton and neutron irradiation studies further show that error modes depend on architecture, memory hierarchy, precision, and mitigation \cite{badia2022proton,wang2021impact}. Candidate devices should therefore be tested with representative models, precisions, compiler settings, memory layouts, and power modes, and the runtime should expect reliability to drift with aging and exposure.

\noindent\textbf{Rad-hard computing trades speed for assurance.}
Radiation-hardened processors, protected memories, ECC, lockstep execution, triple modular redundancy, hardened nonvolatile memory, and scrubbing reduce important failure classes, but often lag COTS devices in performance, efficiency, memory bandwidth, accelerator support, tools, and software ecosystems. TMR and lockstep architectures show how redundancy improves reliability, while LEON-class processors illustrate hardened embedded cores in space computing \cite{iturbe2019tcls,keller2017leon3}. Radiation-tolerant nonvolatile memory and heterogeneous modular redundancy extend this toolbox for storage, checkpointing, and mixed-assurance systems \cite{marinella2021nvm,rogenmoser2025hmr}. The design question is not rad-hard versus COTS in general, but which functions require hardened assurance and which can use COTS acceleration with containment, redundancy, and validation.

\noindent\textbf{Non-maintainability demands autonomous recovery.}
After launch, failed boards cannot normally be replaced, connectors reseated, debuggers attached, or subsystems manually power-cycled, and contact gaps delay ground intervention. Recovery must therefore be onboard. Autonomous recovery work shows how systems can regain functionality after component failures, and microreboots show that fine-grained restart can recover software components without rebooting an entire system \cite{cully2008remus,candea2004microreboot}. Virtual-machine and process checkpointing capture, migrate, or restore execution state after faults \cite{le2011rehype,ansel2009dmtcp}. Satellite AI services should be decomposed into restartable tasks with bounded side effects, persistent checkpoints, watchdogs, safe-mode fallbacks, and policies for partial mission completion.

\noindent\textbf{Fault tolerance spans the full stack.}
A reliable AI service requires more than a reliable accelerator. Radiation or aging can affect weights, activations, compiler output, allocators, drivers, operating-system state, storage metadata, scheduler decisions, and packets. TensorFI exposes neural-network sensitivity through fault injection, and Thales demonstrates algorithm-aware error detection and correction for DNN accelerators \cite{chen2020tensorfi,tyagi2023thales}. Hypervisor replication masks failures below the application, while delay-tolerant networking preserves delivery semantics across disrupted links \cite{bressoud1995hypervisor,fall2003dtn}. Space AI fault tolerance should therefore be layered: ECC and scrubbing for memory, replication or diversity for critical computations, checkpointing for progress, sanity checks for outputs, and disruption-tolerant protocols for data movement.

\noindent\textbf{Runtime must be radiation-aware.}
A radiation-aware runtime should connect environmental exposure, error telemetry, workload criticality, and update trust. Secure update and remote attestation support trusted deployed software, while Ranger and selective hardening show that neural-network resilience can target vulnerable operations rather than harden everything uniformly \cite{asokan2018assured,chen2021ranger,feng2010shoestring}. CLEAR shows how low-cost recovery can support resilient execution under expected transient faults \cite{cheng2016clear}. The runtime should adapt scrubbing rates, checkpoint intervals, model variants, replication, voting, and placement based on orbit-dependent risk and measured faults, and it should support secure post-launch updates for models, kernels, and drivers.

\subsection{Materials, Structures, and Packaging}
\label{subsec:materials-structures-packaging}

\noindent\textbf{Terrestrial servers do not fit orbit.}
Space AI infrastructure cannot be built by moving terrestrial servers into orbit. Data-centre servers assume grid power, active cooling, replaceable parts, benign radiation, continuous networking, human maintenance, and a stationary mechanical environment. Satellite computing measurements show that COTS devices become constrained by power, temperature, software deployment, and platform integration in orbit \cite{xing2024cots}. Constellation systems such as EagleEye and Geoduck show that onboard compute must be co-designed with sensing opportunity, orbital motion, contacts, and mission coordination \cite{cheng2024eagleeye,cheng2025geoduck}. Space--air--ground surveys likewise frame space nodes as constrained edge infrastructure in a heterogeneous network \cite{shen2024survey}. The baseline should therefore be spacecraft-integrated computing, with physical survival, sustained thermal throughput, containment, and validated operating envelopes treated as first-order goals.

\noindent\textbf{Materials determine cooling and lifetime.}
Materials set heat spreading, radiation exposure, contamination risk, structural survival, and lifetime. Graphene- and graphite-based spreaders can reduce local gradients, but their value depends on interface resistance, anisotropy, thickness, manufacturability, cycling stability, and structural compatibility \cite{shahil2012graphene,ji2014graphite}. Shielding reduces particle exposure but adds mass, can generate secondary radiation, and may alter thermal paths, so it must be optimized rather than maximized \cite{durante2011radiation}. Low-outgassing materials are also essential because condensable volatiles can contaminate optics, arrays, coatings, and sensors \cite{pastore2020outgassing}. Thermal interfaces, adhesives, coatings, shields, underfills, boards, and panels should therefore be selected as one coupled materials stack.

\noindent\textbf{Packaging defines the thermal path.}
Heat does not travel from an AI die directly to space; it crosses die attach, package lid, thermal interface, board copper, vias, connectors, fasteners, structural panels, heat straps or pipes, and finally a radiator. Thermal optimization of 3-D integrated circuits shows that placement and inter-layer heat paths shape package-scale hotspots \cite{samal2014thermal}. Liquid-cooling and multiobjective thermal-control research show that cooling capacity, transport cost, and temperature uniformity must be co-optimized \cite{chen2017liquid,sabry2011multiobjective}. At spacecraft scale, radiator panels and heat-transport hardware show that structural layout determines which external surfaces can reject heat \cite{torres2014satellite}. Design tools should propagate workload power maps through package, board, and spacecraft thermal networks rather than collapse the system into a single junction-to-ambient value.

\noindent\textbf{Launch loads constrain hardware form.}
Before an AI module computes, it must survive vibration, acoustic loading, quasi-static acceleration, sine vibration, and separation or pyrotechnic shock. Structural-validation work emphasizes correlating analysis with modal and environmental testing rather than relying on unvalidated finite-element predictions \cite{aglietti2019spacecraft}. Euclid virtual-shaker analysis and flight-like additively manufactured satellite structures illustrate the coupled design--analysis--test loop required for unconventional architectures \cite{pederbelli2023euclid,spicer2022espa}. Random-vibration experiments show that board response, component mass, package location, and solder-joint fatigue can dominate electronics failure risk \cite{alyafawi2009random}. These loads disfavor tall heat sinks, unsupported accelerator cards, socketed modules, long connectors, and airflow-dependent layouts, so mechanical and thermal design must be verified together.

\noindent\textbf{SWaP-C limits compute scaling.}
Mass, volume, power, and cost prevent scaling by simply adding accelerators. Extra compute can require larger arrays, more battery capacity, stronger power conditioning, larger radiators, heat transport, structure, shielding, and qualification. Multidisciplinary small-satellite optimization shows that subsystem and operational variables are tightly coupled, while structural sizing treats mass, stiffness, and performance as competing objectives \cite{hwang2014mdo,ravanbakhsh2012multiobjective}. Constellation sensing studies add orbit design, fleet size, geometry, and communication strategy to the trade space \cite{cheng2024eagleeye}. Efficient accelerators reduce compute, memory, and data-movement cost, but do not remove spacecraft-level tradeoffs \cite{chen2017eyeriss}. The objective is mission value over a Pareto frontier spanning throughput, accuracy, latency, energy reserve, thermal margin, mass, volume, reliability, cost, and software maturity.

\noindent\textbf{Co-design must span chip to spacecraft.}
The physical foundation of space AI infrastructure requires chip--package--board--spacecraft co-design. AI workload traces should inform accelerator architecture, memory hierarchy, package power maps, board placement, power distribution, structural modes, radiation protection, radiator sizing, and orbital thermal boundaries. OpenMDAO illustrates multidisciplinary analysis for coupled models, and spacecraft optimization shows that physical variables and operational schedules can be optimized together \cite{gray2019openmdao,hwang2014mdo}. Cool-3D and optimization-architecture surveys show how architecture, thermal modeling, modularity, convergence, and model ownership can be balanced during early design \cite{cool3d2025,martins2013mdo}. The deliverable should be a validated operating envelope and runtime policy set, not only a bill of materials; sustained, safe, recoverable AI service is the target, and silicon throughput is one input.

\section{Case Studies: From Edge AI to Space-Native Infrastructure}

Section~2.2 surveyed global efforts to deploy AI in orbit. Here, we trade breadth for depth by examining three first-party deployments on BUPT-1, a 12U Sun-synchronous satellite at approximately 490~km, and BUPT-2, a sibling platform with onboard cloud-native computing. Their telemetry, update logs, service-continuity records, and thermally matched 1:1 ground twin enable paired space--ground analysis rarely possible with external deployments.

The three cases form a dependency ladder from node to platform to service: Section~5.2 examines orbital computing capacity, Section~5.3 studies capability deployment and update, and Section~5.4 investigates stateful service continuity under interruption. Together, they reveal the physical, operational, and recovery abstractions required for dependable in-orbit AI systems.

\subsection{The Node: Measuring the Physical Envelope of COTS Compute in Orbit}
\label{sec:case-node}
 
Our first case uses a satellite as an instrument for measuring what an orbital compute node \emph{is}. On a 12U Sun-synchronous smallsat near 490\,km, we flew two Raspberry~Pi~4B boards and two Atlas~200~DK AI accelerators as COTS payloads, logging power, battery state, and structural surface temperature every few seconds over roughly four months. The question is not peak TOPS, but whether capacity can be treated as constant. It cannot. Usable capacity is modulated by two orbital constraints, a thermal ceiling and an energy budget, so the node is better modelled as a \emph{physically modulated resource} than as a fixed server.

\begin{figure}[t]
\centering
\includegraphics[width=\linewidth]{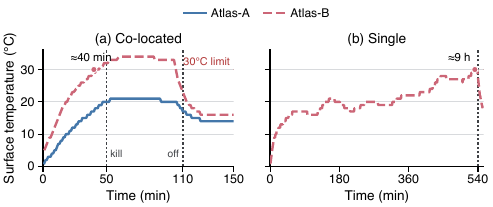}
\Description{Two side-by-side line plots compare accelerator surface temperature.
When Atlas-A and Atlas-B are co-located, Atlas-B reaches the 30 degree Celsius
limit in about 40 minutes; Atlas-A's task is killed near 50 minutes and Atlas-B
powers off near 110 minutes. Operating Atlas-B alone delays the limit to about
nine hours.}
\caption{Thermal ceiling and the co-location effect: a shared radiating surface
collapses a nine-hour single-device window to under two hours.}
\label{fig:node-thermal}
\end{figure}

The thermal dimension is a hard ceiling enforced by the platform. Heat from a COTS chip conducts into a shared aluminium surface and, without convection, leaves only by radiation; $T_{\mathrm{surf}}(t)$ then approaches the 30\,$^{\circ}$C structural limit. One accelerator at full load reaches this limit after about nine hours and stops cleanly. Two co-located accelerators share the same radiating surface: the surface reaches 30\,$^{\circ}$C in roughly forty minutes, the operating system kills the first task near fifty minutes, and the platform powers down the second device at about 110~minutes (Fig.~\ref{fig:node-thermal}). Co-location therefore collapses the safe continuous-compute window from about nine hours to under two, showing that thermal capacity is shared across devices.

\begin{figure}[t]
\centering
\includegraphics[width=\linewidth]{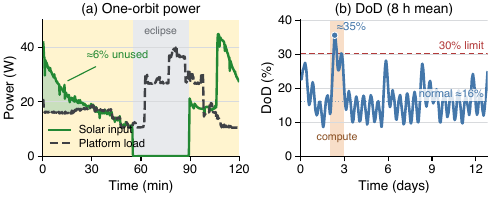}
\Description{The left panel shows solar input and platform load across sunlight
and eclipse, including unused solar power. The right panel shows the eight-hour
mean battery depth-of-discharge, which rises from about 16 percent to 35 percent
during continuous computation and crosses the 30 percent limit.}
\caption{The energy budget and its two governors: solar power in sunlight,
depth-of-discharge in eclipse.}
\label{fig:node-energy}
\end{figure}

The energy dimension is one budget with two governors that switch at the terminator (Fig.~\ref{fig:node-energy}). In sunlight, compute should track harvested solar power from below: exceeding it draws from the battery, while leaving too much converted solar power unused wastes energy and adds heat. About 6\% of converted solar energy goes unused in our measurements. In eclipse, every joule comes from the battery, and the budget is the remaining depth-of-discharge ($\mathrm{DoD}$) headroom. Solar is a perishable flow, whereas $\mathrm{DoD}$ is cumulative and history-dependent. Normal operation keeps daily-average $\mathrm{DoD}$ near 16\%, but multi-day continuous computation drives it to about 35\%, beyond the 30\% design limit; raising the average from 25\% to 30\% costs on the order of a quarter of battery lifetime. Where a task runs in the orbit therefore changes both capacity and mission cost.

Both limits are clocked by orbital mechanics: a thermal ceiling driven by compute heat and unused solar, and an energy budget governed by solar input in sunlight and $\mathrm{DoD}$ in eclipse. In this half-year deployment, computation is bounded by thermal and energy constraints on minute-to-hour timescales; radiation is not the operative limit for this COTS node at this altitude and duration. Capacity also varies within a device, as thermal throttling lengthens inference latency by about 10\% on some models, although this effect is secondary to platform-level task termination.

The missing piece is an abstraction, not telemetry. $T_{\mathrm{surf}}$, $\mathrm{DoD}$, and solar phase are already measured and largely predictable. Yet the compute layer still sees a flat server rather than a forecasted physical envelope. Platform state must be exposed as a schedulable resource, and long-running services must be built to survive involuntary physical preemption.

\textit{Implication:} an orbital node is a physically modulated resource, with capacity metered by surface temperature, solar power, and depth-of-discharge. Like contact windows, this envelope is predictable, but it remains largely invisible to the scheduler that must plan against it.
 
\subsection{The Platform: Application Delivery and Update as the Orbital Control Plane}

\begin{figure}[t]
\centering
\includegraphics[width=0.98\linewidth]{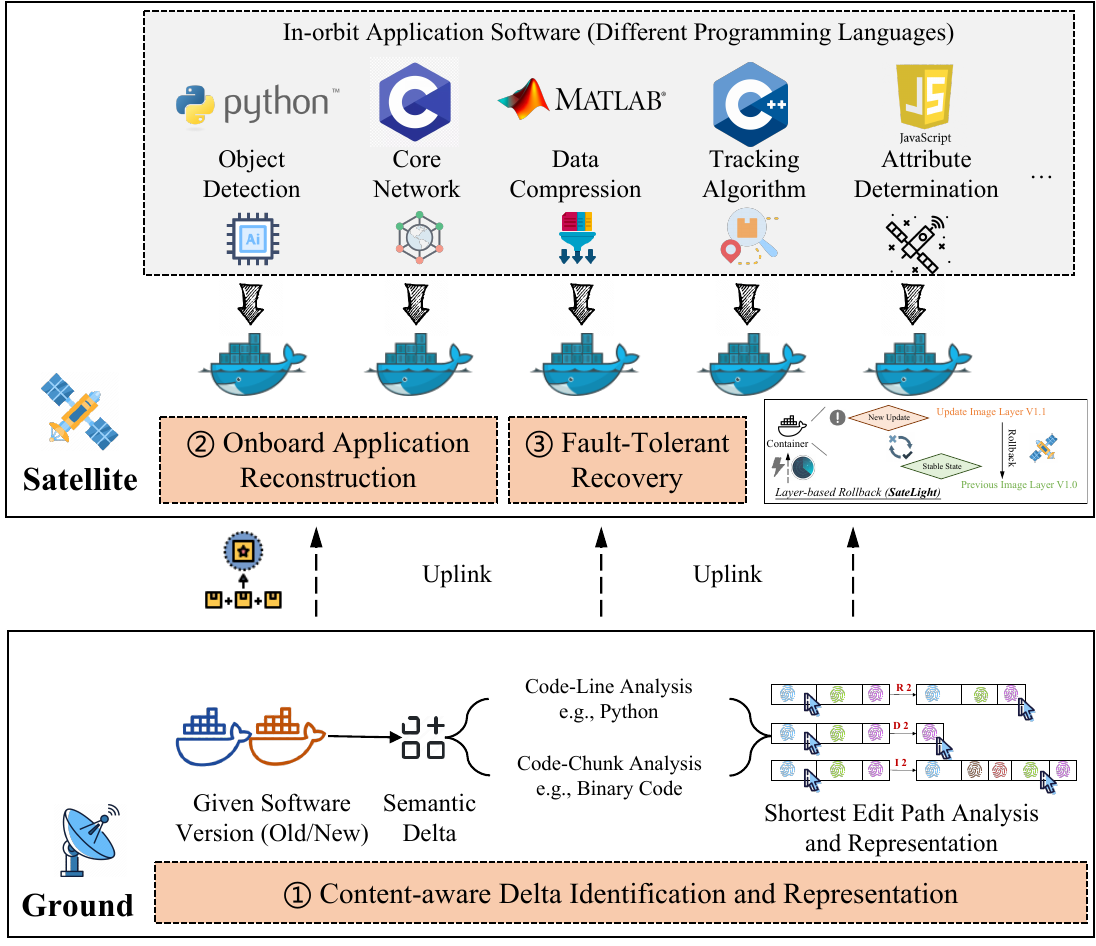}
\caption{The overall workflow of SateLight.}
\label{fig:SateLight}
\end{figure}

To support onboard software maintenance, we propose \textit{SateLight}, a container-based framework for satellite application updates. As shown in Figure~\ref {fig:SateLight}, ground-side tooling compares original and updated containerized applications, extracts content-aware deltas, and encodes semantic changes into compact metadata. The onboard component reconstructs the application through fine-grained updates and uses container layering for fast rollback after failure. We deploy \textit{SateLight} on BUPT-2, an operational LEO satellite with cloud-native onboard computing and ground communication. Across heterogeneous applications, \textit{SateLight} reduces ground-to-satellite transmission latency by up to 91.18\%, with an average reduction of 56.54\%, and achieves 100\% update correctness.

\begin{itemize}
    \item \textbf{Deployment and capability.} SateLight spans ground and satellite with containerized applications, content-aware differential upload, fine-grained onboard reconstruction, and container-layer rollback. It was exercised on ten applications across six languages, from CV inference to 5G core functions, and validated by a live in-orbit update on BUPT-2.
    \item \textbf{Infrastructure reading.} A satellite becomes a platform only when what runs in orbit can be changed safely after launch. Containerization handles application heterogeneity; semantic deltas fit kbps uplinks and ten-minute contacts; layer rollback provides recovery without waiting for the next ground pass.
    \item \textbf{Binding constraints and evidence.} Uplinks run at tens-to-hundreds of kbps with 4--6 ten-minute contacts per day. A full representative vision container costs ${\sim}10.7$ hours of pure uplink time at 200~kbps, roughly two weeks of wall-clock contacts, whereas a content-aware delta for a 10\% code change is tens of kilobytes and ships in seconds. Average transmission-latency reduction is 56.5\% over the best baseline, up to 91\%, with ${\sim}2$~s onboard reconstruction overhead and 100\% correctness. Failed updates roll back autonomously with 0.48~ms backup preparation and ${\sim}36$~s recovery.
    \item \textbf{AI specificity.} Model weights and adapters are among the largest, most frequent artifacts AI infrastructure must ship, and they are binaries. The chunking used for executables can also delta-ship future weight updates. Container layers provide version-level reversibility, while Case~III supplies finer execution-state continuity.
    \item \textbf{Missing abstraction.} SateLight assumes stateless applications and one satellite. Kbps uplinks cannot carry GB-scale weight refreshes even as deltas, pointing to feeder-link capacity, on-orbit adaptation, fleet-scale staged rollout, and update scheduling constrained by energy, thermal state, and contacts.
\end{itemize}
\textit{Implication:} operability precedes intelligence. Without delivery, update, and rollback, an onboard model is a one-shot demo; with them, the satellite becomes an iterable platform.
 
\subsection{The Service: Sustaining Stateful VLM Inference Across Thermal Interruptions}
\label{sec:case-service}

Our third case asks whether a stateful AI service can make forward progress when its compute substrate is repeatedly removed by spacecraft protection mechanisms. We deploy \textit{Rover}, an intermittent-aware VLM inference runtime, on BUPT-1 and BUPT-2 using COTS payloads. Sustained Qwen3-VL-2B inference raises device temperature from 45\,$^{\circ}$C to the 55\,$^{\circ}$C shutdown threshold in only 48~s, while a video-level Earth-observation request can require several minutes of continuous reasoning. A conventional runtime therefore loses visual features, prefill progress, KV caches, and partial outputs whenever the spacecraft bus powers off the payload, and repeated restart from scratch may prevent the request from ever completing.

Rover treats these power losses as recoverable execution boundaries rather than terminal failures. Its checkpoint policy follows the structure of the VLM pipeline. During vision encoding, it persists layer-boundary hidden states because bidirectional attention prevents patches from being resumed independently. During language-model prefill, it partitions execution jointly along token and layer dimensions and checkpoints only the KV tensors and boundary hidden states needed to reconstruct a two-dimensional progress frontier. During decoding, it batches multiple KV-cache updates into each checkpoint, avoiding the fragmented writes and high normal-path overhead of per-token persistence. Phase-specific layouts and committed checkpoint records further ensure that incomplete writes are not mistaken for valid recovery state.

Recovery is also scheduled around model dependencies rather than replayed in a fixed layer-first or chunk-first order. After reboot, Rover restores the weights, KV tensors, and hidden states that unlock useful computation earliest, while overlapping later SSD reads with dependency-ready blocks. This is important because arbitrary shutdowns leave an irregular staircase-shaped frontier: some token--layer blocks have completed, others must be recomputed, and each candidate recovery order exposes a different amount of storage I/O. Rover reduces this restoration problem to a lightweight scheduling plan that can be solved within 100~ms.

\begin{figure}[t]
\centering
\includegraphics[width=0.98\linewidth]{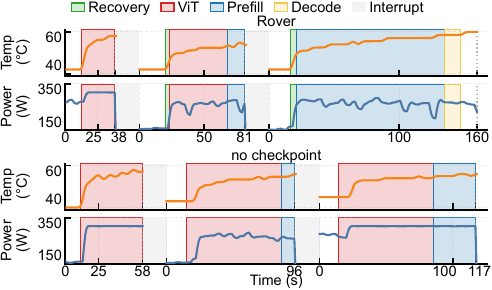}
\caption{Thermal trace for a real-world deployment of Rover.}
\label{fig:rover}
\end{figure}

Experiments driven by in-orbit thermal traces show up to a 3.2$\times$ end-to-end speedup over existing checkpointing baselines, up to an 85.6\% reduction in recovery latency, and up to a 3.56$\times$ reduction in SSD writes. Normal-path energy overhead remains between 3.7\% and 10.4\%. In the real-world satellite deployment, Rover completes most of the requests while the baseline completes none. Figure \ref{fig:rover} shows a representative trace where a request spans three execution windows and two payload resets, but resumes from committed progress and finishes in the third window.

\textit{Implication:} for stateful orbital AI, survivability is an AI-state management problem. A long-running service exists only when its execution state is explicit, durable, and recoverable across physical interruptions.

\subsection{Lessons Learned}
\begin{itemize}
    \item \textbf{The node is not a server.} Capacity is a physics-modulated time series over two measurable variables, enclosure temperature and depth-of-discharge. These envelopes must become scheduler inputs (Case~I; Sections~6.1, 6.3).
    \item \textbf{At LEO, thermal and energy bind before radiation.} Multiple months of unhardened COTS operation produced zero observed SEE errors, while overheating terminated tasks twice. Near-term infrastructure effort should prioritize thermal and energy management (Case~I; Section~4).
    \item \textbf{Operability precedes intelligence.} Without delivery, update, and rollback sized to kbps uplinks and minute-scale windows, each onboard model remains a one-shot demo. Lifecycle support is the entry ticket to infrastructure (Case~II; Sections~6.4, 6.5).
    \item \textbf{Demanding services shift the bottleneck from compute to state.} Container layers provide version-level reversibility, and runtime checkpoints provide execution-level continuity. Both are state management under physics-driven preemption (Cases~II--III; Section~6.4).
    \item \textbf{Predictability generalizes beyond topology.} Energy harvest and thermal accumulation, like contact windows, are forecastable from orbital mechanics. Infrastructure should schedule around forecasts rather than react to thresholds (Cases~I and~III; Sections~6.2, 6.3).
    \item \textbf{Hybrid-first, single-node so far.} Each working system is a space--ground loop, with the ground packaging, validating, and deciding while orbit executes. All three cases remain single-satellite; constellation-scale orchestration is still open (all cases; Section~6 and the roadmap).
\end{itemize}
\textit{Takeaway:} in orbit, communication and computation are both metered by predictable physical cycles. Space-native AI infrastructure is the systems layer that turns those forecasts into scheduling decisions.

\section{Research Challenges: Toward Space-Native AI Infrastructure}

The previous sections show that AI infrastructure in space is a stack-level problem, not a missing component. Compute, networking, model lifecycle, physical state, and mission control must be designed together. Terrestrial AI systems provide abstractions for clusters, accelerators, schedulers, and serving; space systems provide predictable orbits, constrained links, and tightly managed platforms. The challenge is to connect these traditions without importing datacentre assumptions. We identify five problems that stand between isolated onboard demonstrations and space-native infrastructure.

\subsection{Space-Native Infrastructure Abstractions}

A first challenge is the absence of a shared abstraction. Terrestrial AI systems reason about servers, racks, clusters, jobs, services, and regions. Space systems are still usually built around one spacecraft, payload, mission, or pipeline. A model that runs on one satellite therefore does not automatically become portable, schedulable, or maintainable across another satellite, orbital plane, ground station, or mission profile.

The orbital unit is not simply a server. A satellite combines compute, storage, sensors, radios, attitude control, power generation, batteries, thermal paths, fault-management logic, and mission priorities. Its capacity varies with contacts, sunlight, eclipse, thermal exposure, and visibility. A constellation is not a static cluster either, because orbital planes, inter-satellite links, handovers, downlinks, and mission roles determine which nodes can cooperate. Useful abstractions must therefore expose orbit, contact, energy, thermal state, reliability role, storage pressure, and mission priority as scheduling resources.

The boundary is delicate. Exposing every spacecraft detail makes the platform unusable for AI developers; hiding physical state makes the runtime unsafe. A promising direction is a layered resource model in which satellites export forecasted envelopes for compute, power, thermal headroom, storage, and contacts, while the mission system retains authority over safety-critical rules. AI services would request objectives, such as latency, energy budget, degradation mode, and latest downlink time, and the infrastructure would map them to onboard execution, deferral, migration, or ground fallback. This requires resource descriptors, service-level objectives, telemetry schemas, and programming models that are both space-aware and usable by AI system builders.

\subsection{Topology-Predictive Runtime and Cross-Layer Orchestration}

Space networks are dynamic but not arbitrary. Visibility, contact duration, path length, and inter-satellite connectivity are largely driven by orbital mechanics and can be forecast. A runtime with this forecast can pre-position model artifacts, defer nonurgent movement, reserve downlink capacity, migrate service state before handover, and decide whether inference should run onboard, on a neighboring satellite, at a ground station, or in the cloud.

Topology decisions are entangled with compute and mission state. Model placement determines data movement; contacts determine whether checkpoints, adapters, or data products can synchronize; energy and thermal margins determine whether a visible node can accept work; and mission tasks may preempt AI services even when a path exists. A topology-aware runtime must therefore schedule communication together with compute, storage, power, and thermal state.

A central direction is to make contact prediction an operating-system-like primitive. The runtime should expose future topology, contact windows, expected link quality, and uncertainty to placement and admission control. Offline predictions can provide a baseline plan, online telemetry can correct it, and the orchestrator can revise placement, migration, and downlink decisions. Forecasts will still fail because of weather, pointing constraints, congestion, faults, emergency events, and demand bursts. Degradation modes, such as smaller models, batched inference, compressed features, skipped updates, or local-only execution, should therefore be explicit and schedulable.

\subsection{Energy-Thermal-Aware Sustainable AI Serving}

Onboard AI capacity is not a fixed operations-per-second number. It is a time-varying service envelope shaped by sunlight, eclipse, battery depth of discharge, thermal history, heat rejection, payload activity, and safety rules. A spacecraft may run a model briefly but not sustain it as a service. Section~5 shows this directly: thermal protection or energy limits can preempt computation before nominal accelerator capability is exhausted.

This changes AI serving. Datacentre serving systems often optimize throughput, latency, utilization, and cost under stable facility support. In orbit, serving must also preserve energy margin, thermal headroom, and mission safety. A request should be admitted only if the predicted physical trajectory remains safe, and the best placement may be the node whose future contact, battery state, radiator orientation, and mission schedule can support the service without later shutdown.

The research problem is to model service capacity as a forecasted physical resource. The runtime should estimate how many inferences can be served before eclipse, a thermal threshold, or a downlink opportunity; which model variant fits a power cap; how work should move among orbit, ground, and cloud as thermal margin changes; and how urgent mission events should be balanced against background updates or batch analytics. This requires telemetry-driven models of power draw, heat accumulation, cooldown, and workload-dependent accelerator behavior. It also requires safe degradation: smaller batches or models, lower frame rates, delayed tasks, compressed outputs, checkpointing, or transfer to another node or ground backend.

\subsection{AI State, Model Lifecycle, and Distributed Execution}

AI infrastructure manages more than tasks and packets. It manages weights, adapters, prompts, checkpoints, KV caches, intermediate features, sensor data, telemetry, logs, downlink queues, versions, validation reports, and rollback states. In terrestrial clouds these artifacts rely on abundant storage, fast networks, and frequent operator intervention. In space they must survive intermittent links, limited storage, narrow uplinks, radiation faults, thermal interruption, and long lifetimes.

Lifecycle control is the first challenge. Models launched with the spacecraft will become stale as tasks, environments, and ground pipelines evolve. Updates are necessary but risky: they may exceed contact windows, conflict with dependencies, be unsafe under current power conditions, or be impossible to roll back. A space-native model registry should span ground and orbit, tracking versions, dependencies, validation, hardware compatibility, energy and memory profiles, and rollback plans. Deployment should be staged, contact-aware, mission-aware, and verified before activation.

Execution-state continuity is the second challenge. Stateful services carry context that is expensive to reconstruct, including KV caches, partial outputs, intermediate features, and runtime checkpoints. When a node approaches a thermal limit, loses contact, or enters a mission-critical period, the infrastructure must decide whether to checkpoint, migrate, discard, recompute, or degrade state. Distributed execution adds state placement: workloads may split across sensor-side filtering, feature extraction, onboard inference, ground refinement, and cloud aggregation, but split points are useful only if they fit the network and physical envelope. The long-term goal is a lifecycle system in which models and state can be updated, audited, recovered, and composed across space and ground.

\subsection{Trustworthy Evaluation, Resilience, and Governance}

The final challenge is trust. Space AI infrastructure operates inside non-maintainable, high-risk, long-lifetime physical systems. Accuracy, latency, throughput, and model size are necessary but insufficient. A service can be accurate in isolation yet unusable if it consumes too much energy, overheats the platform, misses contacts, cannot recover, or cannot be audited after autonomous action.

Benchmarks should therefore report model quality together with energy margin, thermal headroom, storage pressure, link use, recovery time, update cost, and mission value. They should include realistic orbital traces, contact plans, power and thermal profiles, workload bursts, and failure events. Digital twins can connect simulation, hardware-in-the-loop testing, and in-orbit validation, but only if calibrated with flight telemetry.

Resilience and governance must be built into the infrastructure layer. Radiation events, software faults, model drift, corrupted updates, cyber attacks, link disruption, and physical protection events need explicit handling through authenticated delivery, verified rollback, degraded modes, fault injection, checkpoint recovery, and audit trails. High-impact actions should use verify-before-act mechanisms that separate model recommendation from mission authority when ground supervision is delayed. APIs, logs, update mechanisms, evaluation metrics, and control boundaries should also reflect spectrum, orbital slots, ground-station sharing, cybersecurity, sensed-data privacy, debris risk, and responsible autonomy from the beginning.

\section{Research Roadmap}

The path toward AI infrastructure in space should be staged. Near term, the goal is not a full orbital datacentre but onboard AI that is measurable, updateable, and safe beyond isolated demonstrations. Mid term, satellites, ground stations, and cloud backends should form a hybrid infrastructure loop. Long term, constellations should expose compute, storage, networking, and physical-state resources as programmable services.

\subsection{Near Term: AI-Enabled Satellites}

In the near term, single-satellite AI demonstrations should become reproducible and maintainable systems. Missions should report not only whether a model ran, but also its energy cost, thermal trajectory, memory footprint, contact use, recovery behavior, and operational constraints. These traces and profiles create the empirical basis for schedulers and benchmarks.
Near-term systems should also standardize the minimum lifecycle: package a model, validate it on the ground, deliver it through constrained links, activate it safely, monitor it in orbit, and roll it back if needed. Full autonomy is unnecessary. A hybrid-first design can keep heavy training, validation, and policy decisions on the ground while satellites execute bounded inference, filtering, and local adaptation. The success criterion is deployment, monitoring, update, and evaluation under realistic orbital constraints, not a one-time demonstration.

\subsection{Mid Term: Hybrid Space-Ground AI Infrastructure}

The mid-term stage should connect AI-enabled satellites into hybrid space--ground infrastructure. Satellites act as frontends close to sensors and users, while ground stations and cloud backends provide heavier computation, model management, long-term storage, and fleet-level control. The boundary expands from one spacecraft to the loop of satellite, inter-satellite link, gateway, mission control, model registry, and cloud service.
The defining capability is predictive orchestration. Runtimes should use contact, energy, thermal, storage, and mission forecasts to decide where models and state reside. A model update may be staged before a contact window, a checkpoint moved before a thermal interruption, and a sensing workload run onboard during low-latency events but deferred when contacts make ground processing viable. Mid-term systems should support stateful AI services, contact-aware placement, energy--thermal-aware scheduling, multi-satellite collaboration, calibrated digital twins, and service-level objectives that include physical margins and recovery behavior.

\subsection{Long Term: Orbital AI Infrastructure}

In the long term, AI infrastructure in space could become a shared orbital computing substrate rather than a mission-specific feature. Constellations may expose compute--storage--network resources as programmable infrastructure, multiple services may share orbital nodes, and AI tasks may execute across spacecraft, ground stations, and terrestrial clouds as one distributed system.
This requires constellation-level resource abstraction, bounded autonomy, distributed state management, service migration, secure multi-tenant isolation, and governance-aware scheduling. Orbital contact, energy, thermal headroom, reliability state, and sustainability constraints must be native resources. The goal is not to move all AI computation into space, but to place capability where data, latency, resilience, or mission value justify it and coordinate that capability with terrestrial infrastructure. Progress should be evaluated by whether AI services become sustainable, recoverable, updateable, and composable across the space--ground continuum.

\section{Conclusion}

AI infrastructure in space is emerging as a new systems problem at the intersection of AI systems, networking, and spacecraft engineering. This paper has reviewed the computing foundations, examined the physical constraints that distinguish orbital platforms from terrestrial infrastructure, and used in-orbit case studies to illustrate how these constraints affect computing capacity, software operation, and service continuity. Together, these observations show that space-native AI infrastructure requires resource abstractions and system mechanisms that account for time-varying connectivity, energy, thermal conditions, and recoverability.
Many challenges remain before such infrastructure can support reliable and scalable AI services. Progress will depend on tighter integration across computing, networking, system software, and spacecraft design, as well as realistic evaluation through simulation, testbeds, and in-orbit deployment. We hope this paper helps define a research agenda toward programmable, manageable, and dependable AI infrastructure in space.

\bibliographystyle{unsrt}
\bibliography{references}

\end{document}